\documentclass[12pt]{amsart}

\usepackage{amssymb,amsmath,amscd,color}

\theoremstyle{plain}
\newtheorem{thm}{Theorem}
\newtheorem{theorem}[thm]{Theorem}

\theoremstyle{definition}

\theoremstyle{remark}

\newtheorem*{remark*}{Remark}

\newcommand{\cL}{{\mathcal{L}}}

\newcommand{\cO}{{\mathcal{O}}}

\newcommand{\cU}{{\mathcal{U}}}

\newcommand{\Ai}{{\mathrm{Ai}}}

        \newcommand{\field}[1]{{\mathbb{#1}}}
        \newcommand{\NN}{\field{N}}
        \newcommand{\ZZ}{\field{Z}}
        
        \newcommand{\RR}{\field{R}}
        \newcommand{\CC}{\field{C}}

\allowdisplaybreaks

\begin{document}

\title[Quasi-classical approximation of monopole harmonics]{Quasi-classical approximation of monopole harmonics}
\thanks{This work was supported by the Russian Science Foundation (grant no. 21-41-00018).}

\author[Y. A. Kordyukov]{Yuri A. Kordyukov}
\address{Institute of Mathematics, Ufa Federal Research Centre, Russian Academy of Sciences, 112~Chernyshevsky str., 450008 Ufa, Russia} \email{yurikor@matem.anrb.ru}

\author[I. A. Taimanov]{Iskander A. Taimanov}
\address{Novosibirsk State University, Pirogova st 1, 630090, Novosibirsk; Sobolev Institute of Mathematics, 4 Acad. Koptyug avenue, 630090, Novosibirsk, Russia}
\email{taimanov@math.nsc.ru}

\begin{abstract}
Using the generalization of the multidimensional WKB method to magnetic Laplacians corresponding to monopoles, which we proposed earlier, we obtain explicit formulas for quasi-classical approximations of eigenfunctions for the Dirac monopole.
\end{abstract}

\date{}

\maketitle

\section{Introduction}

In \cite{KT20} (see also \cite{T23}) we proposed a generalization of the multidimensional WKB method (the Maslov canonical operator method \cite{Maslov,MF}) to magnetic Laplacians corresponding to monopoles. This approach can be applied to general differential operators acting on sections of non-trivial bundles.

In \cite{KT20} the quasi-classical parameter $\hbar$ is quantized:
\[
\hbar = 1,\frac{1}{2},\frac{1}{3},\dots,\frac{1}{N},\dots.
\]
For each given value of $\hbar$, almost eigenfunctions of the magnetic Laplacian $\Delta^{L^N}, N = \hbar^{-1}$, corresponding to the magnetic field $NF$ proportional to the initial field $F$, are constructed.

Recently, significant simplifications have been obtained in the computation of the canonical operator near focal points, which have quite simple structure. In particular, they enable to model the canonical operator in standard software packages and effectively apply it to many applied problems \cite{ADNT19,DT22}.

In this paper, we demonstrate a (simultaneous) application of these methods to a test problem, the derivation of explicit formulas for quasi-classical approximations of monopole harmonics, i.e., eigensections, corresponding to the Dirac monopole \cite{Dirac}. Exact solutions for them were obtained in \cite{Tamm,WY}.

\section{Dirac magnetic monopole}

Let $M$ be a Riemannian manifold with a metric $g_{jk}$ and a closed $2$-form $F^{(0)}$ describing the magnetic field. In a sufficiently small neighborhood of each point, we choose a ``vector-potential'' $A^{(0)} = (A^{(0)}_k)$ of this form (a magnetic potential) from the condition
\[
F^{(0)}_{jk} = \frac{\partial A^{(0)}_k}{\partial x^j} - \frac{\partial A^{(0)}_j}{\partial x^k},
\]
where $\{x^j\}$ are local coordinates in this neighborhood. The magnetic potential, generally speaking, is defined only locally.

Consider a particle with charge $q$ on the manifold $M$. In \cite{WY} it is assumed that $q = Ze$, where $Z \in \ZZ$ and $e$ is the elementary charge: $e>0$ and is equal in absolute value to electron charge. Consider  the closed $2$-form on $M$:
\[
F = qF^{(0)}.
\]
If $F$ satisfies the Dirac quantization condition:
\begin{equation}
\label{e:prequantum}
\left[\frac{F}{2\pi}\right] \in H^2(M;\ZZ),
\end{equation}
then the form $F$ is the curvature form of a complex line bundle $\cL$ with structure group $U(1)$, the vector potential $A=qA^{(0)}$ defines the connection $\nabla=d-iA$ in this bundle and the first Chern class of the bundle $\cL$ is equal to
\[
c_1(\cL) = \left[\frac{F}{2\pi}\right].
\]
In this case, one can associate to the magnetic field $F$, or, more precisely, to the system of the external magnetic field $F^{(0)}$ and a particle with charge $q$, the magnetic Laplacian acting on  sections of the bundle $\cL$:
\[
\Delta^\cL = - \frac{1}{\sqrt{g}} \left(\frac{\partial}{\partial x^j} -iA_j\right) \sqrt{g}g^{jk}  
\left(\frac{\partial}{\partial x^k} -iA_k\right).
\]
Here $g = \det \left(g_{jk}\right)$, $g^{jk}g_{km} = \delta^j_m$ and, as usual, summation is assumed over repeated superscripts and subscripts.  

The Dirac monopole \cite{Dirac} describes the motion of an electron in the three-dimensional space in the uniform magnetic field
\[
{\bf H} = g_D \frac{\bf r}{r^3}.
\]
After separating the independent variables into spherical and radial components: $\Psi = f(r)S(\theta,\varphi)$, the stationary Schr\"odinger equation for the eigenfunctions $\Psi(r,\theta,\varphi)$ of the quantized system decomposes into a system of two differential equations. The spherical component $S(\theta,\varphi)$ is a section of some non-trivial line $U(1)$-bundle $\cL$ over the unit two-dimensional sphere with coordinates $\theta$ and $\varphi$. The equation for $S(\theta,\varphi)$ has the form
\[
\Delta^\cL S = E S.
\]

It can be seen from the construction of the connection $A = qA^{(0)}$ and the definition of the magnetic Laplacian that the change of the sign of the charge to the opposite one leads to the conjugation of the bundle $\cL$ and of the eigensections 
\begin{equation}
\label{conjugation}
q \to -q, \ \ \cL \to \bar{\cL} = \cL^\ast, \ \ \psi \to \bar{\psi}, \ \ E \to E, \ \  \mbox{where $\Delta^\cL\psi = E\psi$}.
\end{equation}

In his work, Dirac considered in detail the case when
\[
g_D e = \frac{1}{2}, \ \ \ q = - e,
\]
that is, the magnetic charge $g_D$ of the monopole located at the origin takes the minimum possible positive value so that the quantization condition is satisfied and the particle has the charge $q=-e$ (an electron).
Here we set
$\hbar=c=1$, 
where $\hbar$ is Planck's constant and $c$ is the speed of light in vacuum.
The eigensections corresponding to the smallest eigenvalue $E = \frac{1}{2}$ were found by Tamm \cite{Tamm}:
\begin{equation}
\label{tamm}
S_a = \cos \frac{\theta}{2}, \ \ S_b = \sin \frac{\theta}{2} e^{-i\varphi}
\end{equation}
for $0 \leq \theta < \pi$.

The unit two-dimensional sphere $S^2$ in $\RR^3$ centered at the origin:
\[
S^2=\{(x,y,z)\in \RR^3 : x^2+y^2+z^2=1\},
\]
is endowed with the Riemannian metric induced by the embedding in the Euclidean space $\RR^3$. In spherical coordinates
\[
x=\sin\theta \cos\varphi, \quad y=\sin\theta \sin\varphi, \quad z=\cos\theta, \quad \theta\in (0,\pi), \varphi\in (0,2\pi),
\]
the Riemannian metric $g$ is given by
\[
g=d\theta^2+\sin^2\theta\, d\varphi^2.
\]
Let $F_B$ be the closed 2-form of the magnetic field given by 
\[
 F_B=B\,d \mathrm{vol}=B\sin\theta d\theta\wedge d\varphi,
\]
where $d \mathrm{vol}$ denotes the Riemannian volume form of the metric $g$:
\begin{equation} \label{dvol}
d \mathrm{vol}=\sin\theta d\theta\wedge d\varphi
\end{equation}

The quantization condition \eqref{e:prequantum} means that
\begin{equation}
\label{e:sphere-quant}
2 B\in \ZZ.
\end{equation}

In what follows, we denote by $L$ the bundle with curvature form  
\[
F= \frac 12 \sin \theta d\theta \wedge d\varphi, 
\]
corresponding to the value $B=1/2$. Its first Chern class equals 
\[
c_1(L) = \left[ \frac{F}{2\pi} \right] = 1 \in H^2(S^2;\ZZ) = \ZZ.
\]
In algebraic geometry, it is considered as a bundle over $\CC P^1 = S^2$ and denoted by $L = \cO(1)$. The formula \eqref{tamm} defines sections of its conjugate bundle $\bar{L} = L^{-1} = \cO(-1)$, since it corresponds to a particle with negative charge.   

For $B = \frac{N}{2}$, the form $F_B=NF$ is the curvature form of the line  bundle $L^N$, where $L^N$ is the $N$th tensor power of $L$ for $N>0$, the trivial bundle for $N=0$ and the $N$th tensor power of $\bar{L}=L^\ast$ for $N<0$. It is clear that $c_1(L^N) = N$.
 
Consider the covering of the sphere by two coordinate neighborhoods $U_1$ and $U_2$:
\[
U_1=\left\{0 \leq \theta <\pi, 0\leq \varphi <2\pi\right\}, \quad
U_2=\left\{0 < \theta \leq \pi, 0\leq \varphi <2\pi\right\}.
\]
Since, over each of these domains, the restriction of any line bundle $\cL$ is trivial:
\[
\cL\vert_{U_1} = U_1 \times \CC, \ \ \cL\vert_{U_2} = U_2 \times \CC,
\]
then the bundle is uniquely determined by the clutching function $g_{12}$, which is defined on $U_1 \cap U_2$ and relates the linear coordinates $\lambda$ and $\mu$ in the fibers over $U_1$ and $U_2$:
\begin{equation}
\label{clutch}
\mu = e^{-ic_1(\cL) \varphi} \lambda.
\end{equation}
In particular,  for $\cL = L^N, N=2B$, we have
 \[
 g_{12}(\theta,\varphi) = e^{-2iB\varphi}, \quad (\theta,\varphi)\in U_1\cap U_2.
 \]
 
The magnetic potentials of the form $F_B$ are given by 
\begin{equation}
\label{e:A}
A_1 = B(1-\cos\theta)d\varphi, \quad \text{on}\ U_1;\quad
A_2 = -B(1+\cos\theta)d\varphi, \quad \text{on}\ U_2.
\end{equation}
On $U_1\cap U_2$, they are related by a gauge transformation
\[
A_1 - A_2=g^{-1}_{12}dg_{12}.
\]
The magnetic Laplacian $\Delta^{L^N}$ takes the form
\begin{equation}
\label{e:D}
\begin{aligned}
\Delta^{L^N}= & -\frac{1}{\sin\theta}\frac{\partial}{\partial\theta}\sin\theta\frac{\partial}{\partial\theta}-\frac{1}{\sin^2\theta}\left(\frac{\partial}{\partial\varphi}-iB(1-\cos\theta)\right)^2 \quad \text{on}\ U_1;\\
\Delta^{L^N}= & -\frac{1}{\sin\theta}\frac{\partial}{\partial\theta}\sin\theta\frac{\partial}{\partial\theta}-\frac{1}{\sin^2\theta}\left(\frac{\partial}{\partial\varphi}+iB(1+\cos\theta)\right)^2\quad \text{on}\ U_2,
\end{aligned}
\end{equation}
where $B = \frac{N}{2}$.

The spectrum of the magnetic Laplacian $\Delta^{L^N}$ is computed in
 \cite{Tamm,WY}. It consists of the eigenvalues
\begin{equation}
\label{spectrum}
E_{N,j}=j(j+1)+\frac{N}{2}(2j+1), \quad j=0,1, \ldots,
\end{equation}
of multiplicity
\begin{equation}
\label{multiplicity}
m_{N,j}=N+2j+1.
\end{equation}
The corresponding eigenfunctions are known as monopole harmonics. Explicit formulas for them are given in  \cite{WY}.

Note that the formula \eqref{tamm} presents $S_a$ and $S_b$ as sections of $L^{-1}$ over $U_1$ and according to \eqref{clutch}, over $U_2$, they are given by the sections
\[
S_a = \cos \frac{\theta}{2} e^{i\varphi}, \ \ S_b = \sin \frac{\theta}{2},
\]  
that is, as sections, they have no singularities anywhere.

Let us introduce a Hermitian scalar product on the space of sections of the $U(1)$-bundle $\cL$:
\begin{equation}
\label{hermit}
\langle \phi \vert \psi \rangle = \int_{S^2} \bar{\phi}\psi \, d\, \mathrm{vol}.
\end{equation}
The sections $S_a$ and $S_b$ are orthogonal with respect to \eqref{hermit}: 
$\langle S_a \vert S_b \rangle =0$.

A more detailed discussion of the statements and constructions of \S 2 is given in \cite{T23}.

\section{WKB-approximation for monopole harmonics}

The magnetic geodesic flow is given by the Hamiltonian
\begin{equation}
\label{hamiltonian}
H(\theta, \varphi, p_\theta, p_\varphi)=p_\theta^2+\frac{1}{\sin^2\theta}p^2_\varphi
\end{equation}
with respect to the twisted symplectic form $\Omega$ on $T^*M$:
\[
\Omega=dp_\theta\wedge d\theta+dp_\varphi\wedge d\varphi +B\sin\theta d\theta\wedge d\varphi.
\]
Since the magnetic Laplacian \eqref{e:D} is the quantization of exactly this expression, we omit the usual factor $\frac{1}{2}$ in front of the right hand side in \eqref{hamiltonian}. Thus, the value $E$ of the Hamiltonian has the meaning of twice the kinetic energy of the particle.
 
The corresponding Hamilton equations have the form:
\[
\dot{\theta}=2p_\theta, \quad \dot{\varphi}=\frac{2}{\sin^2\theta}p_\varphi, \quad \dot{p}_\theta=-\frac{2\cos\theta}{\sin^3\theta}p^2_\varphi+B\sin\theta p_\varphi, \quad \dot{p}_\varphi=-B\sin\theta p_\theta.
\]
The magnetic geodesic flow is superintegrable. Therefore, the partition of the phase space into invariant tori is non-unique and determined, for example, by choosing two commuting
first integrals $I_1$ and $I_2$. It is natural to take the Hamiltonian as one of them:
\[
I_1=H(\theta, \varphi, p_\theta, p_\varphi).
\]
As an additional first integral, we take
\[
I_2=p_\varphi-B\cos\theta.
\]
It is easy to check that the function $I_2$ is everywhere defined.

The corresponding invariant tori $\Lambda$ are parameterized by two parameters $E$ and $P$ and given by the equations
\[
p_\theta^2+\frac{1}{\sin^2\theta}p^2_\varphi=E, \quad p_\varphi-B\cos\theta=P.
\]
The torus $\Lambda (E,P)$ is nonempty if and only if  
\[
P^2<E+B^2.
\]

Put
\[
B=\frac{1}{2}.
\] 
The magnetic Laplacian acts on sections of the bundle $L = \cO(1)$.
This corresponds to the positive charge of the particle (in contrast to the Dirac case), i.e. to a positron.

According to \cite{KT20}, quasi-classical eigensections of the operator $\Delta^{L^N}$ are constructed from invariant Lagrangian tori $\Lambda(E,P)$. These tori must satisfy Bohr-Sommerfeld-type quantization conditions, which hold for
\[
E = \lambda_{N,j}=\frac{j(j+1)+\frac{N}{2}(2j+1)+\frac 14}{N^2},\quad j=0,1,2,\ldots.
\]
\[
P=\frac{k}{N}-\frac 12, \quad -j \leq k \leq N+j,
\]
where $\hbar^{-1} = N \in \NN$, and the quantization condition depends on $\hbar$.

Using the connection of the almost eigenvalues $\widehat{E}_{N,j}$ of the operator $\Delta^{L^N}$ with admissible values of the parameter $E$, we obtain
\begin{equation}
\label{e:hat-lambda}
\widehat{E}_{N,j}=\lambda_{N,j}N^{2}=j(j+1)+\frac{N}{2}(2j+1)+\frac 14, \quad j=0,1, \ldots.
\end{equation}
The multiplicity of the almost eigenvalue $\widehat{E}_{N,j}$ equals
\begin{equation}
\label{e:m-lambda}
\widehat{m}_{N,j}=N+2j+1.
\end{equation}
If we compare them with the formulas \eqref{spectrum} and \eqref{multiplicity} for exact
eigenvalues, we see that the formula \eqref{e:hat-lambda} describes the exact eigenvalues of the magnetic Laplacian up to a constant correction $\Delta\lambda_{N,j}=\frac 14$, and the formula \eqref{e:m-lambda} gives the correct answer for multiplicities of these eigenvalues.

The almost eigenfunctions $U_{N,j,k} \in C^\infty(M,L^N)$ of the operator $\Delta^{L^N}$ corresponding to the almost eigenvalue $\widehat{E}_{ N,j}=\lambda_{N,j} N^2$ are given by  
\begin{equation}\label{e:UN}
U_N = U_{N,j,k} =K_{\Lambda(E,P)}^{1/N}u \in C^\infty(M,L^N), \quad
u(\theta,\varphi)\equiv u_0,
\end{equation}
where $K_{\Lambda(E,P)}^{1/N} : C^\infty(\Lambda(E,P)) \to C^\infty(M,L^N)$ is the canonical operator constructed in \cite{KT20}. For brevity, we will omit the indices $j$ and $k$ where possible.

The torus $\Lambda=\Lambda(E,P)$ consists of all $(\theta, \varphi, p_\theta, p_\varphi)$ such that 
\begin{equation}\label{e:ptheta-pvarphi}
\begin{aligned}
p_\theta = P_\theta(\theta):=& \left(E-\frac{1}{\sin^2\theta}\left(P+\frac 12\cos\theta\right)^2\right)^{1/2}=\frac{\left(a+b\cos\theta+c\cos^2\theta\right)^{1/2}}{\sin\theta};\\
p_\varphi= P_\varphi(\theta):= & P+\frac 12\cos\theta,
 \end{aligned}
\end{equation}
where
\[
a=E-P^2, \quad b=-P, \quad c=-\left(E+\frac 14\right),
\]
$\theta\in [\theta_2, \theta_1]$, $\varphi\in [0,2\pi)$ and 
\begin{equation}\label{e:defz}
\theta_j:=\arccos z_j, j=1,2,
\quad
z_{1,2}=\frac{-b\mp \sqrt{\Delta}}{2c}=\frac{-P/2\mp\sqrt{E^2+E(1/4-P^2)}}{E+1/4},
\end{equation}
$z_{1,2}$ are the roots of the quadratic equation $a+bz+cz^2=0$, $\Delta=b^2-4ac>0$.

Singular points of the restriction of the projection $\pi$ to $\Lambda$ are obtained when $\theta=\theta_1$ or $\theta=\theta_2$. Therefore, the singularity cycle $\Sigma(\Lambda)$ consists of two circles
\[
\{ p_\theta=0, p_\varphi=P+\frac 12\cos\theta_j, \theta=\theta_j, \varphi\in [0,2\pi)\}, \quad j=1,2.
\]
The singular points are of the fold type.

The torus $\Lambda$ is the union of two nonsingular canonical charts $\cU^\pm$ with coordinates
\begin{equation}\label{e:coordU}
\sigma_{\pm} : (\theta,\varphi)\in U:=\left\{\theta\in [\theta_2, \theta_1], \varphi\in [0,2\pi)\right\} \mapsto (\theta,\varphi, \pm P_\theta(\theta), P_\varphi(\theta))\in \cU^\pm\subset \Lambda.
\end{equation}
and the singularity cycle $\Sigma(\Lambda)$:
\[
\Lambda=\cU^+\cup \cU^-\cup\Sigma(\Lambda).
\]
The image of the torus $\Lambda=\Lambda(E,P)$ under the canonical projection $\pi :T^*S^2\to S^2$ has the form
\[
\pi(\Lambda)=\overline{U}=\{(\theta,\varphi) : \theta\in [\theta_2, \theta_1], \varphi\in [0,2\pi)\}.
\]

Recall (see \cite[Formula (22)]{KT20}) that if $\cU$ is a non-singular chart on the Lagrangian manifold $\Lambda$ with coordinate mapping $\sigma : U\to \cU$, then the associated canonical operator $K^h_{\Lambda,\cU} : C^\infty(\cU)\to C^\infty(U)$ is given by
\begin{equation}\label{e:Kh-Lambda}
K^h_{\Lambda,\cU} u(x)=e^{(i/h)\tau(\sigma (x))-\pi i m_{\cU}/2}\sqrt{\left|\frac{d\mu(\sigma(x))}{d\,\mathrm{vol}}\right|} u(\sigma(x)), \quad x\in U,
\end{equation}
where $\tau$ is the eikonal function, $m_{\cU}$ is the Maslov index of the chart $\cU$, $\mu$ is a smooth positive density on $\Lambda$ invariant under the magnetic geodesic flow, $d\,\mathrm{vol}$ is the Riemannian volume form. Since the bundle $L$ is nontrivial, all our constructions also depend on the choice of the trivialization of the bundle $L$, on which the form of the eikonal function $\tau$ depends. As a consequence, for $h=1/N$, the results of applying the operator $K^h_{\Lambda,\cU}$ to a function $u$ on $\Lambda$ for different choices of a chart $\cU$ and of a trivialization of the bundle $L$ give rise to a well-defined section of the bundle $L^N$.

In the case under consideration, the projection $\pi(\cU^\pm)=U$ of the non-singular canonical chart $(\cU^\pm, \sigma_{\pm})$ to $\Lambda$ is contained in two different trivializing neighborhoods $U_1$ and $U_2$. The corresponding expressions for the magnetic potential are given by formulas \eqref{e:A} with $B=\frac 12$. Depending on the choice of trivialization, we obtain different expressions for the eikonal function $\tau^\pm_{U_1}$ and $\tau^\pm_{U_2}$ on $\cU^\pm$ obtained in \cite{KT20}.
If we consider $\pi(\cU^\pm)$ as a subset of $U_1$, then
\[
\tau^\pm_{U_1}(\sigma_{\pm}(\theta,\varphi))=\pm I(\theta)+(P+\frac 12)\varphi+\tau_1,  \quad (\theta,\varphi)\in U.
\]
If we consider $\pi(\cU^\pm)$ as a subset of $U_2$, then
\[
\tau^\pm_{U_2}(\sigma_{\pm}(\theta,\varphi))=\pm I(\theta)+(P-\frac 12)\varphi+\tau_2,  \quad (\theta,\varphi)\in U.
\]
Here the function $I(\theta)$ is given by
\[
\begin{aligned}
I(\theta)= &\frac 12 |P+\frac 12|\arcsin \frac{2a+b+(b+2c)\cos\theta}{(\cos\theta-1)\sqrt{\Delta}} +\frac 12 |P-\frac 12|\arcsin \frac{2a-b+(b-2c)\cos\theta}{(\cos\theta+1)\sqrt{\Delta}}\\
& +\sqrt{E+\frac 14}\arcsin \frac{2c\cos\theta+b}{\sqrt{\Delta}},
\end{aligned}
\]
and $\tau_1$, $\tau_2$ are arbitrary constants.

Let us find a measure $\mu$ on the invariant torus $\Lambda (E,P)$ that is invariant under the magnetic geodesic flow. Let us write it in the chart $\cU^\pm$ in the form
\[
\mu_\pm=f_\pm(\theta,\varphi)\,d\theta \, d\varphi. 
\]
The restriction of the generator of the magnetic geodesic flow to the torus $\Lambda (E,P)$ has the form 
\[
X=\pm2P_\theta(\theta)\frac{\partial}{\partial \theta}+\frac{2}{\sin^2\theta}P_\varphi(\theta) \frac{\partial}{\partial \varphi},
\]
where $P_\theta$ and $P_\varphi$ are given by \eqref{e:ptheta-pvarphi}. 

The condition for the measure $\mu$ to be invariant with respect to the flow is given by 
\[
\pm\frac{\partial}{\partial \theta}(2P_\theta(\theta) f_\pm(\theta,\varphi))+\frac{\partial}{\partial \varphi}(\frac{2}{\sin^2\theta}P_\varphi(\theta) f(\theta,\varphi))=0. 
\]
Setting $f_\pm(\theta,\varphi)=f_\pm(\theta)$, we get
\[
\frac{\partial}{\partial \theta}(2P_\theta(\theta) f_\pm(\theta,\varphi))=0. 
\]
Therefore, as a solution, one can take
\[
 f_\pm(\theta,\varphi)=\frac{1}{P_\theta(\theta)}.
\]
Thus, the invariant measure $\mu$  can be taken to be  the measure
\[
\mu_\pm=\frac{\sin\theta}{\left(a+b\cos\theta+c\cos^2\theta\right)^{1/2}} \,d\theta\, d\varphi. 
\]
It is easy to check that $\mu$ is a smooth positive density on the torus $\Lambda$.

Recall that the Riemannian volume form $d\, \mathrm{vol}$ is given by \eqref{dvol}.
 
Let us compute the indices of charts. First of all, we recall that the cotangent bundle $T^*S^2$ is endowed with the twisted symplectic form $\Omega$. Therefore, in order to apply   well-known formulas for the Maslov index, it is necessary to pass to the standard symplectic form $\Omega_0$. Such a passage is given by the choice of a trivialization of the line bundle $L$ and, accordingly, by the choice of a magnetic potential. Considering the trivialization over $U_1$, we obtain the mapping
\[
f_{U_1}(\theta,\varphi,p_\theta,p_\varphi)=(\theta,\varphi,p_\theta,p_\varphi+\frac 12(1-\cos\theta)),
\]
which takes $(T^*S^2\left|_{U_1}\right.,\Omega)$ to $(T^*S^2\left|_{U_1}\right.,\Omega_0)$. Under this mapping, the invariant torus $\Lambda (E,P)$ is mapped to the torus $\Lambda_0(E,P)$ given by the equations
\[
p_\theta^2+\frac{1}{\sin^2\theta}(p_\varphi-\frac 12(1-\cos\theta))^2=E, \quad p_\varphi=P+\frac 12,
\]
and the canonical charts $\cU^\pm$ to nonsingular canonical charts $\cU^\pm_0$ with coordinates 
\[
\sigma_{\pm,0} : (\theta,\varphi)\in U \mapsto (\theta,\varphi, \pm P_\theta(\theta), P+\frac 12)\in \cU_0^\pm\subset \Lambda_0.
\]

Consider the Jacobian 
\[
\mathcal J (\sigma_{\pm}(\theta,\varphi))= f^{-1}(\sigma_{\pm}(\theta,\varphi))=P_\theta(\theta),\quad (\theta,\varphi)\in \cU^\pm.
\]
Choose a central point $\sigma_{+,0}(\theta_+,\varphi_+)\in \cU^+_0$, say, $\sigma_{+,0}(\theta_+,\varphi_+)=\sigma_{+,0}(\pi/2,0)$ and a point $\sigma_{-,0}(\theta_-,\varphi_-)\in \cU^-_0$,  say, $\sigma_{-,0}(\theta_-,\varphi_-)=\sigma_{0,-}(\pi/2,0)$. Choose and fix the argument of the Jacobian at the central point: ${\rm arg}\, \mathcal J (\sigma_{0,+}(\theta_+,\varphi_+))=0$. Accordingly, the index of charts  $\cU^+_0$ and $\cU^+$ equals zero:
\[
m_{\cU^+_0}=m_{\cU^+}=0.
\]

The index of the chart $\cU^-_0$ is given by
\[
m_{\cU^-_0}=\frac{1}{\pi}{\rm arg}\, \mathcal J (\sigma_{+.0}(\theta_+,\varphi_+)) +\operatorname{ind}\gamma,
\]
where $\gamma : [0, 1] \to \Lambda_0$ is a smooth path such that $\gamma(0) = \sigma_{+,0}(\theta_+,\varphi_+)$ and $\gamma(1) = \sigma_{-,0}(\theta_-,\varphi_-)$ and $\operatorname{ind}\gamma$ is the Maslov index of $\gamma$.

The following formula holds (see, for instance, \cite[Formulas (1.1) and (1.2)]{DNS17}):
\[
\operatorname{ind}\gamma =\lim_{\varepsilon\to +0}\underset{\gamma}{\operatorname{var}}\operatorname{arg} \mathcal J^{\varepsilon},
\]
where, for any $\varepsilon>0$, the Jacobian $\mathcal J^{\varepsilon}$ has the form
\[
\mathcal J^{\varepsilon} (\sigma_{\pm}(\theta,\varphi))=P_\theta(\theta)\left(1\mp i\varepsilon\frac{\partial P_\theta}{\partial\theta}\right),\quad (\theta,\varphi)\in \cU^\pm.
\]
Using this formula, it is easy to compute that
\[
\operatorname{ind}\gamma_-=-1,\quad m_{\cU^-_0}=m_{\cU^-}=-1.
\]

By \eqref{e:Kh-Lambda}, in the domain $\left\{(\theta,\varphi) : \theta_2+\varepsilon < \theta <\theta_1-\varepsilon, 0\leq \varphi <2\pi\right\}$ with an arbitrary $\varepsilon>0$, we have
\begin{equation}\label{e:Kh-Lambda-non1}
K^{h,1}_{\Lambda,\cU} u(\theta,\varphi)=\frac{u_+(\sigma_+(\theta,\varphi))e^{\frac{i}{h}\tau^+_{U_1}(\sigma_+(\theta,\varphi))}}{\left(a+b\cos\theta+c\cos^2\theta\right)^{1/4}}+i\frac{u_-(\sigma_-(\theta,\varphi)) e^{\frac{i}{h}\tau^-_{U_1}(\sigma_-(\theta,\varphi))}}{\left(a+b\cos\theta+c\cos^2\theta\right)^{1/4}},
\end{equation}
where we consider $\pi(\cU^\pm)$ as a subset of $U_1$ and $u_\pm$ denotes the restrictions of $u$ to $\cU^\pm$.

The formula \eqref{e:Kh-Lambda-non1} can be rewritten in the form
\begin{multline}\label{e:Kh-Lambda-non2}
K^{h,1}_{\Lambda,\cU} u(\theta,\varphi)=e^{\frac{3\pi i}{4}}e^{\frac{i(\tau^+_{U_1}+\tau^-_{U_1})}{2h}}\Bigg[ \frac{u_--u_+}{\left(a+b\cos\theta+c\cos^2\theta\right)^{1/4}}\cos\left(\frac{\tau^+_{U_1}-\tau^-_{U_1}}{2h}+\frac{\pi}{4}\right)\\
-i\frac{u_-+u_+}{\left(a+b\cos\theta+c\cos^2\theta\right)^{1/4}}\sin\left(\frac{\tau^+_{U_1}-\tau^-_{U_1}}{2h}+\frac{\pi}{4}\right)\Bigg].
\end{multline}
Set
\[
\tau^\pm_1=\mp I(\theta_2).
\]
Then 
\[
\tau^+_{U_1}+\tau^-_{U_1}=2(P+\frac 12)\varphi, \quad \tau^+_{U_1}-\tau^-_{U_1}=2(I(\theta)-I(\theta_2)),
\]
and the formula \eqref{e:Kh-Lambda-non2} takes the form
\begin{multline}\label{e:Kh-Lambda-non3}
K^{h,1}_{\Lambda,\cU} u(\theta,\varphi)=e^{\frac{3\pi i}{4}}e^{\frac{i(P+1/2)\varphi}{h}}\Bigg[\frac{u_--u_+}{\left(a+b\cos\theta+c\cos^2\theta\right)^{1/4}}\cos\left(\frac{I(\theta)-I(\theta_2)}{h}+\frac{\pi}{4}\right) \\
-i\frac{u_-+u_+}{\left(a+b\cos\theta+c\cos^2\theta\right)^{1/4}}\sin\left(\frac{I(\theta)-I(\theta_2)}{h}+\frac{\pi}{4}\right)\Bigg].
\end{multline}

In particular, setting $u(\theta,\varphi)\equiv u_0$, we obtain a formula for the almost eigenfunction $U_N \in C^\infty(M,L^N)$ of the operator $\Delta^{L^N} $ with the corresponding eigenvalue $\hat\lambda_N$
\begin{equation}\label{e:UN1}
U^{(1)}_N(\theta,\varphi)=
\frac{2e^{\frac{\pi i}{4}}u_0}{\left(a+b\cos\theta+c\cos^2\theta\right)^{1/4}}e^{i(P+1/2)N\varphi}\sin\left((I(\theta)-I(\theta_2))N+\frac{\pi}{4}\right),
\end{equation}
where we consider $\pi(\cU^\pm)$ as a subset of $U_1$. 

Similarly, we obtain
\begin{equation}\label{e:UN2}
U^{(2)}_N(\theta,\varphi)=
\frac{2e^{\frac{\pi i}{4}}u_0}{\left(a+b\cos\theta+c\cos^2\theta\right)^{1/4}}e^{i(P-1/2)N\varphi}\sin\left((I(\theta)-I(\theta_2))N+\frac{\pi}{4}\right),
\end{equation}
where we consider $\pi(\cU^\pm)$ as a subset of $U_2$. 

Formulas \eqref{e:UN1} and \eqref{e:UN2} are valid in the domain  
$\left\{(\theta,\varphi) : \theta_2+\varepsilon < \theta <\theta_1-\varepsilon, 0\leq \varphi <2\pi\right\}$ with an arbitrary $\varepsilon>0$. 

We use the method proposed in \cite{ADNT19} to describe the canonical operator in a neighborhood of a fold of a Lagrangian manifold.

Let us write the function $U^{(1)}_N(\theta,\varphi)$ in terms of the Airy function $\Ai$, using the well-known asymptotic representation for $\Ai$ in the form of trigonometric functions:
\[
\Ai(-x)= \frac{1}{\sqrt{\pi}x^{1/4}}\sin\left(\frac 23x^{3/2}+\frac{\pi}{4}\right)(1+O(x^{-1})), \quad x\to +\infty. 
\]
We get
\[
U^{(1)}_N(\theta,\varphi)\asymp
\sqrt{\pi}e^{\frac{\pi i}{4}} e^{i(P+1/2)N\varphi}A(\theta)N^{1/6} \Ai\left(-N^{2/3}\Phi(\theta)\right),
\]
where $\asymp$ means the equality up to terms of order $O(1/N)$ in the amplitude, 
\[
\Phi(\theta)=\left(\frac{3}{2}(I(\theta)-I(\theta_2))\right)^{2/3}, 
\]
\[
A(\theta)=\frac{2u_0}{\left(a+b\cos\theta+c\cos^2\theta\right)^{1/4}}\left(\frac{3}{2}(I(\theta)-I(\theta_2))\right)^{1/6}.
\] 

Let us find the asymptotics of the amplitude and phase in the formula as $\theta \to \theta_{2}+0$ in order to extend these asymptotic formulas to some neighborhood of $(\theta_2-\varepsilon, \theta_2]$. Note that for $\theta \to \theta_{2}+0$ 
\[
a+b\cos\theta+c\cos^2\theta=c(\cos\theta-\cos\theta_1)(\cos\theta-\cos\theta_2)\sim c(\cos\theta_1-\cos\theta_2)\sin\theta_2 (\theta-\theta_2),
\]
\[
\frac{(a+b\cos\theta+c\cos^2\theta)^{1/2}}{\sin\theta}\sim \frac{(c(\cos\theta_1-\cos\theta_2))^{1/2}}{(\sin\theta_2)^{1/2}}(\theta-\theta_2)^{1/2},
\]
and, therefore,
\begin{align*}
I(\theta)-I(\theta_2) & =\int_{\theta_2}^\theta \frac{(a+b\cos\theta+c\cos^2\theta)^{1/2}}{\sin\theta}d\theta \\
& \asymp \int_{\theta_2}^\theta \frac{(c(\cos\theta_1-\cos\theta_2))^{1/2}}{(\sin\theta_2)^{1/2}}(\theta-\theta_2)^{1/2}d\theta \\ & = \frac{2}{3}\frac{(c(\cos\theta_2-\cos\theta_1))^{1/2}}{(\sin\theta_2)^{1/2}} (\theta-\theta_2)^{3/2}.
\end{align*}
We get the asymptotics of the phase for $\theta \to \theta_{2}+0$ 
 \[
\Phi(\theta)\asymp \frac{(c(\cos\theta_2-\cos\theta_1))^{1/3}}{(\sin\theta_2)^{1/3}} (\theta-\theta_2)
\]
and of the amplitude 
 \begin{align*}
A(\theta) \asymp & \frac{2u_0}{(c(\cos\theta_1-\cos\theta_2)\sin\theta_2 (\theta-\theta_2))^{1/4}}\frac{(c(\cos\theta_2-\cos\theta_1))^{1/12}}{(\sin\theta_2)^{1/12}} (\theta-\theta_2)^{1/4}\\ = & \frac{2u_0}{(c(\cos\theta_1-\cos\theta_2))^{1/6}(\sin\theta_2)^{1/3}}.  
\end{align*}
 
From  \eqref{e:defz}, we find 
\[
\sin \theta_{1,2}=\frac{P\sqrt{E} \mp 1/2\sqrt{E+1/4-P^2}}{E+1/4}
\]
and
\[
c(\cos\theta_1-\cos\theta_2)=\sqrt{\Delta}=\sqrt{E^2+E(1/4-P^2)}.
\]

Thus, we see that the functions $\Phi(\theta)$ and $A(\theta)$ are smooth on the half-interval $[\theta_2, \theta_2+\varepsilon)$, including the point $\theta_2$, moreover $\Phi^\prime(\theta_2)\neq 0$. We extend them to the half-interval $(\theta_2-\varepsilon, \theta_2)$, using the above asymptotics for $\theta \to \theta_{2}+0$.
 
Using arguments similar to those given in the proof of Theorem 2 in \cite{ADNT19}, we obtain the following statement.

\begin{theorem}
The function $U^{(1)}_{N,j,k} (\theta,\varphi)$ given by \eqref{e:UN} has the form 
\[
U^{(1)}_N(\theta,\varphi)\asymp \sqrt{\pi}e^{\frac{\pi i}{4}} e^{i(P+1/2)N\varphi}A(\theta)N^{1/6} \Ai\left(-\left(\frac{3}{2}N\right)^{2/3}\Phi(\theta)\right),
\]
in the domain $\left\{(\theta,\varphi) : \theta_2-\varepsilon < \theta <\theta_1-\varepsilon, 0\leq \varphi <2\pi\right\}$ with some $\varepsilon>0$, where we consider  $\pi(\cU^\pm)$ as a subset of $U_1$ and
\[
\Phi(\theta)=\begin{cases} (I(\theta)-I(\theta_2))^{2/3}, & \text{if}\ \theta\in [\theta_2, \theta_2+\varepsilon)\\ \frac 23 \frac{(c(\cos\theta_2-\cos\theta_1))^{1/3}}{(\sin\theta_2)^{1/3}} (\theta-\theta_2)
, & \text{if}\ \theta\in (\theta_2-\varepsilon, \theta_2) \\
\end{cases}
\]
\[
A(\theta)=\begin{cases} \frac{2u_0}{\left(a+b\cos\theta+c\cos^2\theta\right)^{1/4}}\left(\frac{3}{2}(I(\theta)-I(\theta_2))\right)^{1/6}, & \text{if}\ \theta\in [\theta_2, \theta_2+\varepsilon)\\ \frac{2u_0}{(c(\cos\theta_1-\cos\theta_2))^{1/6}(\sin\theta_2)^{1/3}}
, & \text{if}\ \theta\in (\theta_2-\varepsilon, \theta_2) \\
\end{cases}
\]
\end{theorem}
 
For $N=1, j=0, k=1,2$ we obtain quasi-classical approximations of the eigensections
$\bar{S}_a$ and $\bar{S}_b$ of the form \eqref{tamm} (recall that we are considering the case of a particle with a different sign of the charge, which leads to the conjugation of bundles and their sections).
In general, a comparison of these quasi-classical answers with the classical ones (see Appendix) shows that they are almost localized on the projections of the corresponding Lagrangian tori. 
 
It is easy to see that for any $N$ the sections $U_{N,j,k}, j=0,1,2,\ldots, -j \leq k \leq N+j,$ are almost orthogonal with respect to the scalar product \eqref {hermit}:
\[
\langle U_{N,j_1,k_1} \vert U_{N,j_2,k_2} \rangle \asymp 0, \quad (j_1,k_1)\neq (j_2,k_2). 
\]
Indeed, for different $j$, they are almost orthogonal, since they are almost eigensections corresponding to different almost eigenvalues of the operator $\Delta^{L^N}$, while, for the same $j$ and different $k$, they are almost orthogonal, because they have the form $S(\theta)e^{ik\varphi}$.

Moreover, if we choose the constant $u_0$ so that the normalization condition holds true:
\[
\int_{\Lambda (E,P)}u^2_0\,d\mu=1\Leftrightarrow u_0=\frac{1}{\sqrt{\mu(\Lambda (E,P))}}=\frac{1}{2\pi}(E+1/4)^{-1/4},
\]
then the corresponding almost eigensection $U_{N,j,k}$ is asymptotically normalized: $$\|U_{N,j,k}\|\asymp 1.$$

\section{Monopole harmonics}
 
By \cite{WY}, for the eigenvalue $E_{N,j}, j=0,1, \ldots,$ of $\Delta^{L^N}$ given by \eqref{spectrum}, the corresponding eigensections $Y_{N,j,k}, k=-j,\ldots,N+j,$ are of the form
\[
Y^{(1)}_{N,j,k}(\theta,\varphi)=\Theta_{N,j,k}(\theta)e^{ik\varphi} \quad \text{on}\ U_1; \quad 
Y^{(2)}_{N,j,k}(\theta,\varphi)=\Theta_{N,j,k}(\theta)e^{i(k-N)\varphi} \quad \text{on}\ U_2,
\]
where the function $\Theta_{N,j,k}(\theta)$ has the form  
\[
\Theta_{N,j,k}(\theta)=\widetilde\Theta_{N,j,k}(\cos\theta),
\]
where
\[
\widetilde\Theta_{N,j,k}(x)=\left[\frac{N+2j+1}{4\pi}\frac{(N+j+k)!(j+k)!}{(N+j)!j!}\right]^{1/2}\left(\frac{1-x}{2}\right)^{-k/2}\left(\frac{1+x}{2}\right)^{(N-k)/2}\widetilde P_{N,j,k}(x),
\]
the polynomial $\widetilde P_{N,j,k}(x)$ is the Jacobi polynomial $P^{\alpha,\beta}_n(x)$ with parameters
\[
\alpha=-k, \quad \beta=N-k, \quad n=j+k, 
\]
and has the form
\[
\widetilde P_{N,j,k}(x)=j!(j+N)!\sum_{s}\frac{(-1)^{j+k-s}}{s!(j-s)!(N-k+s)!(j+k-s)!}\left(\frac{1-x}{2}\right)^{j+k-s}\left(\frac{1+x}{2}\right)^{s},
\]
where the sum is taken over all integers $s$ for which the arguments of the factorials are non-negative.

It is easy to see that
\[
Y^{(1)}_{N,j,k}(\theta,\varphi)=e^{iN\varphi}Y^{(2)}_{N,j,k}(\theta,\varphi).   
\]

The sections $Y_{N,j,k}$ are eigenfunctions of the angular momentum operator $L_z$ given by
\[
L_zY^{(1)}= (-i\partial_\varphi - \frac N2) Y^{(1)} \quad \text{on}\ U_1; \quad L_zY^{(2)}= (-i\partial_\varphi + \frac N2) Y^{(2)}  \quad \text{on}\ U_2.
\]
It is easy to see that
\[
L_z Y_{N,j,k} = (k-\frac N2) Y_{N,j,k}.
\]

In particular, for $N=1, j=0$, we have
\[
\widetilde \Theta_{(1,0,0)}(x)=\frac{1}{\sqrt{2\pi}} \left(\frac{1+x}{2}\right)^{1/2},\quad \Theta_{(1,0,0)}(\theta)=\frac{1}{\sqrt{2\pi}} \cos \left({\theta}/{2}\right)
\]
\[
Y^{(1)}_{(1,0,0)}(\theta,\varphi)=\frac{1}{\sqrt{2\pi}} \cos \frac{\theta}{2}, \quad 
Y^{(2)}_{(1,0,0)}(\theta,\varphi)=\frac{1}{\sqrt{2\pi}} \cos \frac{\theta}{2} e^{-i\varphi}.
\]
and
\[
\widetilde \Theta_{(1,0,1)}(x)=-\frac{1}{\sqrt{2\pi}} \left(\frac{1-x}{2}\right)^{1/2},\quad \Theta_{(1,0,1)}(\theta)=-\frac{1}{\sqrt{2\pi}} \sin \left({\theta}/{2}\right)
\]
\[
Y^{(1)}_{(1,0,1)}(\theta,\varphi)=-\frac{1}{\sqrt{2\pi}} \sin \frac{\theta}{2} e^{i\varphi}, \quad 
Y^{(2)}_{(1,0,1)}(\theta,\varphi)=-\frac{1}{\sqrt{2\pi}} \sin \frac{\theta}{2}.
\]
We see that $Y^{(1)}_{(1,0,0)}(\theta,\varphi)$ and $Y^{(1)}_{(1,0,1)}( \theta,\varphi)$ coincide with the solutions \eqref{tamm} up to normalization (division by $\sqrt{2\pi}$) and conjugation. According to \eqref{conjugation}, the conjugation is due to the fact that
these solutions describe particles with charges $e$ and $-e$ in the same magnetic field. 
 
 \vskip5mm
 
The authors thank S.Yu. Dobrokhotov, who drew their attention to this problem and to the papers \cite{ADNT19,DT22}, where a simplification of the formulas for the Maslov canonical operator near the simplest singularities of Lagrangian manifolds is proposed.


\begin{thebibliography}{00}
\bibitem{KT20}
Yu. A. Kordyukov, I. A. Taimanov, Quasi-classical approximation for magnetic monopoles. Uspekhi Mat. Nauk 75 (2020), no. 6(456), 85--106 [Russian Math. Surveys 75 (2020), no. 6, 1067--1088]

\bibitem{T23}
I. A. Taimanov,
Geometry and quasi-clasical quantization of magnetic monopoles,
Teoret. Mat. Fiz. (to appear)

\bibitem{Maslov}
V. P. Maslov, Perturbation theory and asymptotic methods, Moscow State Univ, Moscow, 1965

\bibitem{MF}
V. P. Maslov, M. V. Fedoryuk, Semi-Classical Approximation for Equations of Quantum Mechanics, Nauka, Moscow, 1976; Semi-Classical Approximation in Quantum Mechanics, Reidel, Dordrecht, 1981.


\bibitem{ADNT19}
A. Yu. Anikin, S. Yu. Dobrokhotov, V. E. Nazaikinskii,  A. V. Tsvetkova, Uniform asymptotics in the form of an Airy function for semiclassical bound states in one-dimensional and radially symmetric problems. Teoret. Mat. Fiz. 201 (2019), no. 3, 382--414 [Theoret. and Math. Phys. 201 (2019), no. 3, 1742--1770]

\bibitem{DT22}
S.Y. Dobrokhotov, A.A. Tolchennikov, 
Keplerian Trajectories and an Asymptotic Solution of the Schr\"odinger Equation with Repulsive Coulomb Potential and Localized Right-Hand Side. Russ. J. Math. Phys. 29 (2022), 456--466. 

\bibitem{Dirac}
P. A. M. Dirac, Quantised singularities in the electromagnetic field, Proc. Roy. Soc. London Ser. A, 133 (1931), 60--72.

\bibitem{Tamm}
Ig. Tamm, Die verallgemeinerten Kugelfunktionen und die Wellenfunktionen eines Elektrons im Felde eines Magnetpoles, Z. Phys., 1931, no. 3--4, 141--150.

\bibitem{WY}
T.T. Wu, C.N. Yang,
Dirac monopole without strings: monopole harmonics, Nuclear Phys. B, 107:3 (1976), 365--380.

\bibitem{DNS17}
S. Yu. Dobrokhotov, V. E. Nazaikinskii, A. I. Shafarevich,  New integral representations of Maslov's canonical operator in singular charts. Izv. Ross. Akad. Nauk Ser. Mat. 81 (2017), no. 2, 53--96 [Izv. Math. 81 (2017), no. 2, 286--328]
\end{thebibliography}
\end{document}